# Enhanced Water Nucleation and Growth Based on Microdroplet Mobility on Lubricant-Infused Surfaces

*Jianxing Sun[a], Xinyu Jiang[a], and Patricia B. Weisensee[a,b,*]*

a. Department of Mechanical Engineering & Materials Science, Washington University in St. Louis, USA

b. Institute of Materials Science and Engineering, Washington University in St. Louis, USA

* Corresponding author: p.weisensee@wustl.edu

**Abstract** Lubricant-infused surfaces (LISs) can promote stable dropwise condensation and improve heat transfer rates due to a low nucleation free-energy barrier and high droplet mobility. Recent studies showed that oil menisci surrounding condensate microdroplets form distinct oil-rich and oil-poor regions. These topographical differences in the oil surface cause water microdroplets to rigorously self-propel long distances, continuously redistributing the oil film and potentially refreshing the surface for re-nucleation. However, the dynamic interplay between oil film redistribution, microdroplet self-propulsion, and droplet nucleation and growth is not yet understood. Using high-speed microscopy, we reveal that during water condensation on LISs, the smallest visible droplets (diameter ~ 1μm, qualitatively representing nucleation) predominantly emerge in oil-poor regions due to a smaller thermal activation barrier. Considering the significant heat transfer performance of microdroplets (< 10 μm) and transient characteristic of microdroplet movement, we compare the apparent nucleation rate density and water collection rate for LISs with oils of different viscosity and a solid hydrophobic surface at a wide range of subcooling temperatures. Generally, the lowest lubricant viscosity leads to the highest nucleation rate density. We characterize the length and frequency of microdroplet movement and attribute the nucleation enhancement primarily to higher droplet mobility and surface refreshing frequency. Interestingly and unexpectedly, hydrophobic surfaces outperform high-viscosity LISs at high subcooling temperatures, but are generally inferior to any of the tested LISs at low temperature differences. To explain the observed non-linearity between LISs and the solid hydrophobic surface, we introduce two dominant regimes that influence the condensation efficiency: mobility-limited and coalescence-limited. We compare these regimes based on droplet growth rates and water collection rates on the different surfaces. Our findings advance the understanding of dynamic water-lubricant interactions and provide new design rationales for choosing surfaces for enhanced dropwise condensation and water collection efficiencies.

**Keywords**  Lubricant-infused surface; LIS; SLIPS; Dropwise condensation; Nucleation rate

# Introduction

Achieving stable dropwise condensation is desirable in many industrial applications, such as water-harvesting,[1,2] desalination,[3] power generation,[4] and thermal management[5] due to higher heat transfer performance and water collection rates as compared to traditional filmwise condensation. In recent decades, many efforts have been dedicated to creating various non-wettable engineered surfaces,[6] such as superhydrophobic,[7,8] hybrid,[9] and lubricant-infused surfaces (LISs)[10–12] that enable the formation of discrete condensate droplets and enhance water shedding from the surface once condensed.[13] Notably, LISs can promote stable dropwise condensation of water and low surface energy liquids.[12,14] Onset of heterogeneous water nucleation intrinsically occurs more easily on a soft or liquid surface than on a solid one under the same thermal conditions.[15–17] On LISs, droplets nucleate at the lubricant-vapor interface due to a low nucleation barrier, and later move into the lubricant due to cloaking and capillary forces.[18,19] After growing to a certain size and protruding the lubricant film, droplets can shed from the surface at relatively small critical diameters under gravity due to an extremely low contact angle hysteresis.[20] Although the oil layer on LISs can cause an additional thermal resistance compared to solid hydrophobic coatings, LISs have a superior heat transfer performance compared to filmwise condensation or dropwise condensation on thin hydrophobic coatings up to an oil thickness of ~ 50-80 μm.[21] This significantly enhanced heat transfer performance of LISs has been attributed mainly to the higher mobility and lower nucleation barrier.[22,23]

It is well known that on LISs oil pulls up along the sides of droplets to form oil menisci due to capillary forces.[24–27] On the macroscale, the size of the oil menisci surrounding a millimetric droplet is very small compared to the size of the droplet. However, we previously showed that oil menisci surrounding microdroplets on nanostructure-based LISs are profound and of similar size as the droplets themselves.[28] These oil menisci are large enough to effectively create oil-rich regions surrounding microdroplets or microdroplet clusters. Due to conservation of mass, oil-poor regions consequently form in the droplet-empty spaces, creating a microscopically uneven oil film during condensation. Recent studies have reported that microdroplets in oil-poor regions and the transition areas between oil-poor and oil-rich regions are able to robustly self-propel long distances and thereby coalesce with neighboring droplets, frequently refreshing the surface along their trajectories.[23,28,29] One question that naturally arises is "how strongly does the microdroplet self-propulsion influence droplet nucleation and growth (and ultimately condensation rates) on LISs?" Answering this question is non-trivial, as these different metrics are tightly coupled: The non-uniform oil film distribution is expected to influence nucleation and droplet mobility, whereas droplet dynamics influence the distribution of the oil film. Despite the importance of microdroplets on condensation heat transfer rates (droplets smaller than 10 μm account for approximately 75% of the total heat transfer), the interplay between nucleation, droplet growth, microdroplet mobility, and lubricant film dynamics remains largely elusive. Macroscopically, the distribution of droplet sizes on LISs has been reported to be independent of lubricant viscosity for steady-state condensation operation,[21] however, lower viscosity lubricant-covered smooth surfaces showed a higher condensation heat transfer coefficient than those with oil of higher viscosity.[30] To better evaluate condensation and heat transfer rates, we propose that it is necessary to examine the transient formation rate of emerging droplets (correlated to the nucleation rate density) over longer periods of time, as opposed to the commonly used time-averaged approach to determine droplet size distributions.[21] The nucleation rate density (NRD) takes into account the area availability for re-nucleation (*i.e.*, a spatial component) and kinetics of nucleation (*i.e.*, a temporal component). Microdroplet self-propulsion is heavily dependent on the lubricant viscosity, with droplets self-propelling faster for lower oil viscosity.[28] Anand *et al.* also observed that the size distribution of condensate droplets at early times and the area fraction of droplet coverage are highly dependent on lubricant viscosity.[18] Here, we hypothesize that a (viscosity-dependent) higher mobility of microdroplets could efficiently refresh the surface at a higher frequency, leading to higher condensate formation and growth rates.

It is impossible to visualize freshly nucleated droplets (size ~ 1-10 nm), and most experimental work interprets nucleation phenomena based on the observation of microscale condensate droplets.[22,31,32] Here, we also focus on the formation and growth of the smallest visible droplets (diameter ~ 1 μm) during water condensation on LISs with oils of different viscosity and solid hydrophobic surfaces in a custom-designed environmental chamber. In the following, we refer to these smallest visible microdroplets as "nuclei", since 1) coalescence is typically thought of as becoming important only for microscopic droplets larger than 1-5 μm[33] and 2) nanodroplets account for a negligible amount of heat towards the overall heat transfer and are hence not relevant for the purpose of this work. The spatial preference of nucleation is first investigated by combining high-speed imaging and optical microscopy. A combination of thermodynamic analysis of the nucleation energy barrier and infrared (IR) thermal imaging of the oil-vapor interface on a LIS during condensation was performed to interpret the nucleation preferences. We present the temporal evolution of the (apparent) nucleation rate density in a complete condensation cycle (nucleation, growth, coalescence, removal/sweeping, back to nucleation) on LISs and hydrophobic surfaces. By conducting a statistical analysis over many such cycles, we present an interdependence between nucleation rate, lubricant viscosity, and the vapor-substrate temperature difference. We also characterized the mobility and the growth rate of microdroplets to better understand the role of droplet self-propulsion and droplet coalescence on condensation rates.

Finally, we performed water collection experiments on LISs of different oil viscosity and a solid hydrophobic surface at different vapor temperatures to support our findings and hypotheses. This work will advance our understanding of the in-plane spatial preference for droplet nucleation on LISs and the significant role of microdroplet mobility on the apparent nucleation rate density as well as microdroplet growth. It will also help inform the selection of a lubricant for LISs to achieve higher heat transfer rates and water collection efficiencies at given temperature differences.

## Experimental Section

**Preparation of lubricant-infused surfaces**

The substrate was obtained by cutting a plain microscope glass slide (Thermo Scientific) into pieces (2.5 cm × 2.5 cm), which were then rinsed with acetone, isopropanol, and de-ionized (DI) water in sequence, and dried with compressed $N_2$. Subsequently, the cleaned glass surface was sprayed with Glaco Mirror Coat Zero (soft 99 Co., Japan) and then placed in the fume hood at room temperature for at least 1 hour. Glaco is a commercially available superhydrophobic agent consisting of the suspension of functionalized silica nanoparticles (~ 40 nm) in alcohol. This process renders a layer of relatively uniform porous nanostructure with a thickness of ~ 1 μm on the surface, as shown in the top view and cross-sectional scanning electron microscope (SEM) images in **Figure 1a-c**. The contact angle of DI water on the Glaco-coated glass slide is 165 ± 3°. In this work, Krytox GPL oils were chosen as the infused liquid, featuring a large range of viscosities at similar surface tensions of $\gamma_{ov}$ = 17 ± 1 mN/m against air (vapor) and $\gamma_{do}$ = 53 mN/m against the water droplets.[27] To obtain the lubricant-infused surface, the nanostructured surface was impregnated with Krytox oils *via* spin coating. Based on the spin coating model developed by Emslie *et al.*,[34] the film thickness scales as $d_{oil} \sim (\mu/t\omega^2)^{1/2}$, where $\mu$ is the oil viscosity, $t$ is the total rotating time of the sample, and *ω* is the angular velocity of the sample. The revolutions per minute (RPM) were set for different viscosity oils as shown in **Table 1**. The initial oil film thickness was measured using high-speed interferometry, which shows good prediction accuracy of the model (**Supplemental Information, Section S1**). The Krytox oils completely wetted the Glaco-coated surface. For a 1.2 mm (diameter) water droplet, the apparent contact angle on these Krytox-infused LISs is approximately 90° with negligible contact angle hysteresis (*i.e.*, advancing and receding apparent contact angles are within the measurement uncertainty of ± 2°). The sliding angles of water droplets on LISs with Krytox 102 and 106 are 2.5° and 5°, respectively. For comparison, hydrophobic samples were also prepared by spin coating the cleaned glass with a layer of 3% Teflon™ AF in FC-40 (Chemours). The hydrophobic surfaces displayed advancing and receding contact angles of 126 ± 3° and 111 ± 3°, respectively.

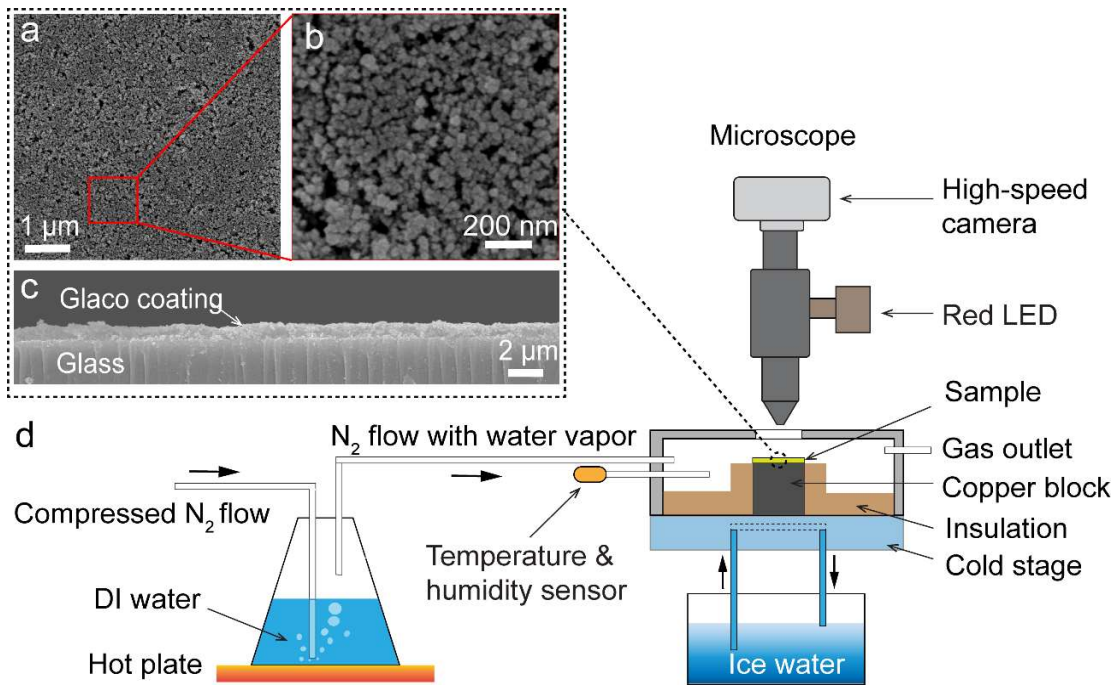

*Figure 1* Experimental set-up and characterization of nanostructured surface before lubricant impregnation. (a-c) Scanning electron microscope (SEM) images of the Glaco superhydrophobic coating prior to oil infusion from top and cross-sectional views. (d) Schematic of the experimental condensation apparatus.

*Table 1* Physical properties at room temperature and corresponding spin coating parameters. The subscript "ov" for the surface tension stands for the oil-vapor interface.

| Lubricant | Viscosity $v$ [cP] | Surface tension $\gamma_{ov}$ [mN/m] | Thermal conductivity [W/m · K] | RPM | Time [s] |
|---|---|---|---|---|---|
| Krytox GPL102 | 73 | 17±1 [27] | 0.08 – 0.09 | 600 | 120 |
| Krytox GPL104 | 350 | 17±1 [27] | 0.08 – 0.09 | 1160 | 120 |
| Krytox GPL106 | 1627 | 17±1 [27] | 0.08 – 0.09 | 2700 | 120 |
| Mineral oil | 148 | 20 [35] | N/A | 800 | 120 |
| DI water | 1 | 72 | 0.6 | N/A | N/A |

*Note:* The viscosity and thermal conductivity information of Krytox oils are obtained from the manufacturer

**Water vapor condensation experiments**

We conducted water vapor condensation experiments on LISs with varying lubricant viscosity at a series of vapor-substrate temperature differences, where $N_2$ served as carrier gas. As shown in **Figure 1d**, a LIS sample was placed on a copper block inside a custom-made chamber. A cold plate connected to the copper block was maintained at a constant temperature ($T_s \approx 276 \pm 1.5$ K) by the circulation of ice water. A flask containing DI water was heated on a hot plate (Thermo Fisher Scientific). Compressed nitrogen gas was supplied to the bottom of the flask at a flow rate of 12.5 ± 1 liters per minute (LPM), controlled by a flowmeter (NFM-TT, Turbo Torch), where it saturated with vapor. A lower flow rate, 4 ± 1 LPM, was set to obtain slow condensation for thermal imaging measurements. The $N_2$-vapor mixture was then guided into the chamber through insulated tubes. The chamber might contain other non-condensable components of air, such as $O_2$ and $CO_2$, but their concentrations are expected to be negligible. The degrading heat transfer performance of dropwise condensation with non-condensable gases (NCG) is entirely attributed to the existence of the diffusion layer and its resistance of the vapor–NCG boundary layer, irrespective of gas composition.[36] The temperature and relative humidity (*RH*) in the chamber were monitored by a temperature and *RH* probe (PCMini52, MICHELL Instruments) with an accuracy of ±1K and ±1% at *RH* = 20% - 90%, ±5% at *RH* > 90%, respectively. The relative humidity value was close to 100% for all experiments. The condensation experiments were monitored under an upright DIY Cerna Microscope (Thorlabs) equipped with brightfield objectives (10×, 50×, 100× L Plan SLWD). An image of the experimental setup is shown in **Section S2** of the **Supplemental Information**. Videos were captured using a Photron FASTCAM Mini AX200 high-speed camera at 60 – 1000 frames per second (fps). For the investigation of the time evolution of the apparent nucleation rate on LISs, the chamber was oriented vertically and we focused on the center of the sample. The sample surface was naturally swept by large sliding droplets due to gravity. In the experiments for studying the average apparent nucleation rate density, the chamber was placed horizontally to circumvent the location-dependence of vertical surfaces. Once we observed the occurrence of large condensate droplets with a diameter of over 700 μm (spanning about two times the field-of-view under the 50 × objective, 379 × 379 μm), we stopped the data collection and then gently tilted the chamber to a vertical position to let droplets shed, leaving a fresh area for the next cycle of nucleation, growth, coalescence, and droplet shedding ("sweeping").

**Statistical analysis of the apparent nucleation rate density (NRD)**

New samples were used for each experiment with a certain combination of lubricant viscosity and vapor temperature, and every experiment was repeated at least three times. We recorded three artificial sweeping cycles for each sample with a 50× objective lens and observed 5–7 locations in each sweep. A sweeping cycle is the time interval from the formation of distinct oil-rich and oil-poor regions to the moment that condensate droplets grow to a certain size and are removed from the surfaces (if vertically placed they naturally shed due to gravity). At each location, a video was recorded with a duration of about 12 seconds and three randomly selected one-second long segments were extracted from each video. We manually counted the number of the smallest emerging visible droplets in the entire field of view, *i.e.*, in both oil-poor and oil-rich regions (and regions covered with droplets) at every frame (interval $\Delta t$). The apparent nucleation rate density was then calculated by $\text{NRD} = \frac{\text{\#new droplets}}{A_{\text{tot}} \cdot \Delta t}$, where $A_{\text{tot}}$ is the field of view of 900 × 950 pixels, that is, 0.117 mm$^2$ for the 50× objective. We used the same approach to determine the NRD on the hydrophobic sample. Finally, we averaged the results from every data point within different sweeping cycles. The smallest detectable droplet with the 50× objective is 1.48 ± 0.37 μm in diameter. Using the 100× objective lens, we were able to detect droplets with approximately 0.74 μm in diameter and observed that droplets smaller than 2 μm rarely coalesce with neighboring droplets. Therefore, it is safe to use the 50× objective to increase the field of view at the expense of resolution and approximate the number of the smallest microdroplets to be representative of the number of nucleating droplets. To a certain extent, our results will underestimate the real nucleation rate density, however, are well suited to elucidate the general effect of microdroplet mobility on nucleation and condensation dynamics.

## Results and Discussion

### Microdroplet self-propulsion on an unevenly distributed oil film of LISs

During condensation on LISs, oil menisci form around numerous microdroplets that protrude from the oil film. On a macroscopic scale, the ratio of the size of the oil menisci versus that of millimetric or submillimetric droplets is fairly small, so the oil film appears to be uniform, as shown in the schematic in **Figure 2a**. However, for microdroplets, the menisci are significant and can lead to a microscopically uneven distribution of the oil film (**Figure 2b**). Oil-rich regions exist surrounding large microdroplets (50 – 500 um), and oil-poor regions naturally appear between the microdroplets. We used an inverted scanning confocal fluorescence microscope (LSM 880 Airyscan, Carl Zeiss) to visualize the uneven distribution of the oil film, where Nile red (Sigma-Aldrich) dyed mineral oil (Hydrobrite 380 PO, Sonneborn) was used with a 488 nm laser beam. As shown in **Figure 2c-d**, oil-rich regions with roughly twice the lateral extension of the central microdroplets commonly exist. We use mineral oil for these measurements due to the unavailability of suitable fluorescent dyes for Krytox oils. Although Krytox oil cloaks water droplets, *i.e.*, forms a thin (~ 10s nm) conformal oil layer surrounding the droplets, whereas mineral oil does not, we expect the oil menisci sizes to be similar. For example, Wu and Miljkovic showed that oil menisci are very comparable for silicon oil (cloaking) and carnation mineral oil (non-cloaking).[37] We hence expect the confocal images of the mineral oil to be a good approximation of the shape and size of Krytox menisci.

Previous studies showed that newly condensed microdroplets ($D < 30$ μm) spontaneously move long distances towards the oil-rich regions due to an unbalanced capillary force.[28,29] As the main, centered microdroplets ($D > 50$ μm) move across the surface and coalesce with neighboring ones, the oil film continuously redistributes. Compared to relatively stationary condensate microdroplets on solid hydrophobic surfaces, the robust microdroplet self-propulsion on LISs potentially increases the chances of droplet coalescence and frequency of refreshing surfaces, which is likely to influence droplet nucleation, growth and thereby the overall condensation rate. In the following sections, we will elucidate the interplay between the oil-redistribution, microdroplet mobility, and nucleation.

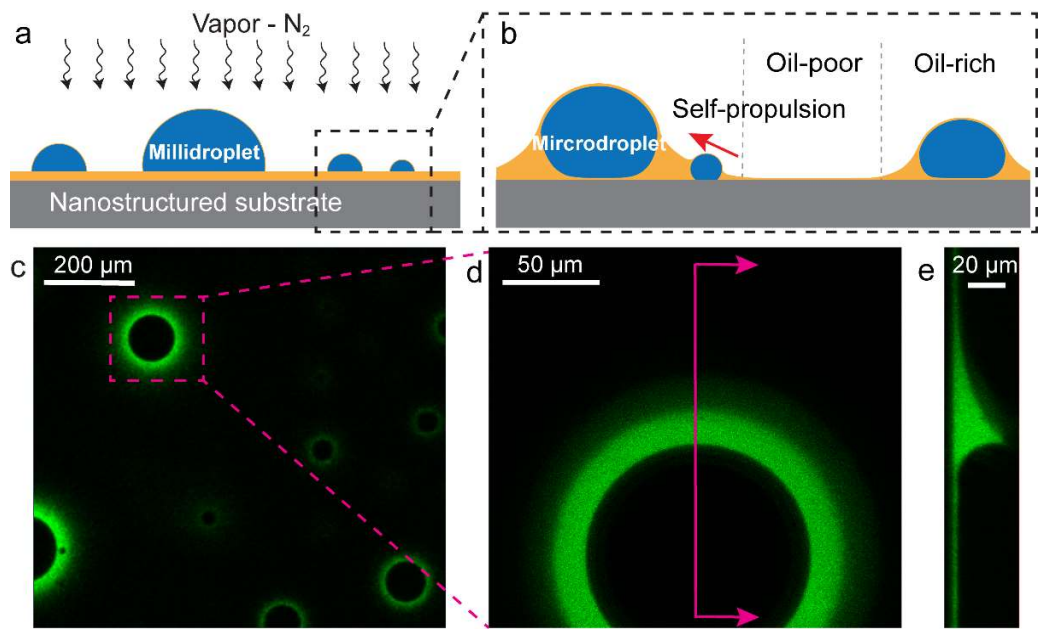

*Figure 2 Microscopically uneven oil film during condensation on lubricant-infused surfaces. A seemingly uniform oil film at the macroscale (a) is in reality unevenly distributed at the microscale (b). Significant oil menisci exist surrounding microdroplets, leading to the formation of oil-rich and oil-poor regions. Small microdroplets spontaneously slide towards oil-rich regions. (c) & (d) show confocal microscope images of water condensation using 10× air and 40× oil immersion objectives, respectively. (e) is the cross-sectional view of a z-stack image from (d), exemplifying a common geometry of the oil meniscus in oil-rich regions (here: Mineral oil dyed with Nile red displays as green).*

**Preferred areas for nucleation: Oil-poor regions**

Water nucleation has been assumed to preferentially initiate at the oil-vapor interface of LISs due to limited vapor diffusion through the oil film as well as the low nucleation energy barrier at the oil-vapor interface.[15,18,38] A recent study showed that satellite droplets can also form on lubricant-cloaked water droplets during condensation on lubricant-infused micro- or nanotextured superhydrophobic surfaces.[39] Here, we focus on the influence of the microscopically uneven oil film on the in-plane spatial preference for nucleation and note that we observe nucleation on top of larger oil-covered droplets only for the highest vapor temperatures. To qualitatively investigate the interplay of nucleation and the dynamic oil film during condensation, we conducted condensation experiments on a LIS with Krytox GPL 104. In the optical image sequences of **Figures 3a**, oil-poor regions (marked with red ellipses) are characterized by having a brighter color, as less light is absorbed and scattered by the thin oil layer compared to a thicker one, and wide interference fringes. Again, we refer to the smallest visible droplets (diameter ~ 1 μm) as "nuclei".[33] We clarify that the number of apparent nuclei might not reflect the real nucleation density, but is sufficient to demonstrate the spatial preference of nucleation. At $t$ = 0.24 s, the image shows that nucleation predominantly occurs in oil-poor regions. As droplets grow and self-propel across the surface, they coalesce with neighboring droplets to create fresh oil-poor regions. **Figure 3b** presents a relatively larger view of a characteristic condensation process using a 20× objective. We observe that nucleation (circled in yellow by ImageJ) is again mainly confined to the oil-poor regions. It is worthy to note that nucleation also occurs to a limited extent at the oil-vapor interface in oil-rich regions or on top of condensed droplets when the vapor-substrate temperature difference is sufficiently high.

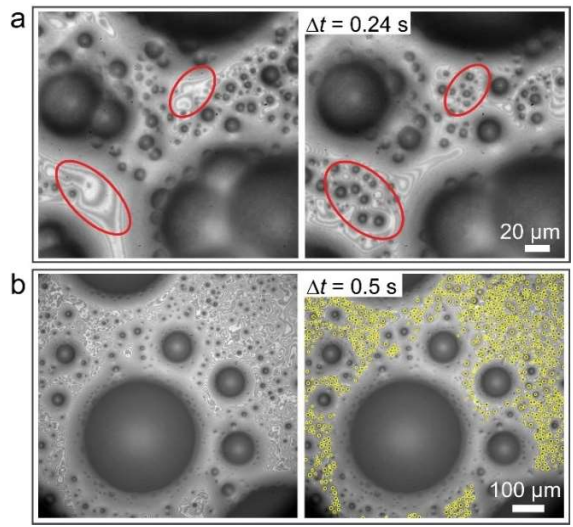

*Figure 3* Experimental image sequences showing preferred nucleation in oil-poor regions. Optical image sequences of emerging droplets on a LIS with Krytox GPL 104 at a magnification of (a) 50× and (b) 20×. Oil-poor regions are marked with red ellipses in (a). Yellow circles in (b) highlight emerging droplets (= nuclei).

We presume that several factors influence the observed spatial preference of nucleation. A higher conduction resistance $R_{oil}$ of the oil film and hence a higher oil-vapor interfacial temperature are expected in the oil-rich (meniscus) regions surrounding the droplets, potentially limiting nucleation in these regions. Along with the high conduction resistance, latent heat released from condensing vapor on existing droplets or the oil menisci will increase the interfacial temperature, making these locations less favorable for nucleation. At the same time, condensing droplets require a continuous radial influx of vapor from the surroundings to compensate for the loss of vapor (*i.e.*, condensation), creating a vapor concentration (or partial pressure) gradient near the droplet surface.[40] This gradient can suppress cluster formation (*i.e.*, nucleation) or growth especially in the close proximity of larger droplets due to an insufficient vapor pressure. It has been previously observed that during condensation in the presence of NGCs, annular nucleation-free zones, so called dry-zones, surround induvial droplets.[41–44] The dry-zone width was previously reported to be 15 – 20 µm at a subcooling of 35 K.[42] With increasing environmental vapor concentration, *i.e.*, increasing gas (vapor) temperature or subcooling, the width of nucleation-free zones decreases.[41] For our work, the question arises: Is the higher interfacial temperature of the oil meniscus or the dry-zone the dominant factor influencing the spatial preference for nucleation, or are they equally important?

In a first step, we compared the lateral extensions of menisci ($l_m$), typical dry-zone dimensions ($\delta_{dz}$), and the occurrence of nucleation. The detailed analysis can be found in **Section S3** of the **Supplemental Information**. When oil menisci surrounding droplets are significantly small ($l_m \sim$ 10 µm) and subcooling is significant enough to create dry-zone, for example 21 K, such that $l_m < \delta_{dz}$, we observe that nucleation predominantly occurs at the boundary between the oil-rich meniscus and the oil-poor region, *i.e.*, within the dry-zone. Similarly, when $l_m > \delta_{dz}$, for example at a subcooling of 41 K and for large menisci ($l_\mathrm{m} \sim 33$ µm), we still observe nucleation at the boundary between oil-rich and oil-poor regions; well outside the dry-zone. Hence, we conclude that the oil film thickness determines the nucleation preference. To substantiate our claim further, in the following, we combine a basic heat transfer analysis in the oil film and the classical nucleation theory to qualitatively show that oil-poor regions are preferred for in-plane nucleation.

In a first-order approach, we approximate the heat transfer through the oil as one-dimensional conduction from the oil-vapor interface (temperature $T_\mathrm{ov}$) to the colder oil-solid interface (temperature $T_\mathrm{s}$):

$$q'' = k_\mathrm{oil} \cdot \frac{T_\mathrm{ov} - T_\mathrm{s}}{d_\mathrm{oil}}, \qquad (1)$$

where $q''$ is the heat flux (assumed to be spatially constant, which is appropriate for the purpose of this work, but should be used with caution when conducting a full-scale heat transfer analysis[45]), $k_\mathrm{oil}$ is the thermal conductivity of the oil, and $d_\mathrm{oil}$ is the local thickness of the oil. For dropwise condensation of steam at near-atmospheric pressure, past work showed that the average heat flux is typically $q'' \sim 100$ to $1500$ kW/m².[46] Here, we make a conservative assumption of a heat flux of $q'' = 100$ kW/m² and $T_\mathrm{s} = 276$ K. **Figure 4b** shows the change of $T_\mathrm{ov}$ with varying $d_\mathrm{oil}$. We notice that the vapor-facing surface temperature is up to 3.5 K higher in oil-rich regions than in oil-poor regions. Corresponding to different temperatures $T_\mathrm{ov}$, the change of the free-energy barrier $\Delta G$ versus the oil film thickness $d_\mathrm{oil}$ (from oil-poor regions to oil-rich regions on LISs) is shown as well. Details on the calculation of $\Delta G$ is available in **Section S4** of the **Supplemental Information**. A bulk vapor temperature of $T_\mathrm{v} = 303$ K was presumed in the calculation. We see that the energy barrier for nucleation is approximately one order of magnitude lower in oil-poor regions than in oil-rich regions, which confirms our hypothesis and initial observations that nucleation predominantly occurs in oil-poor regions.

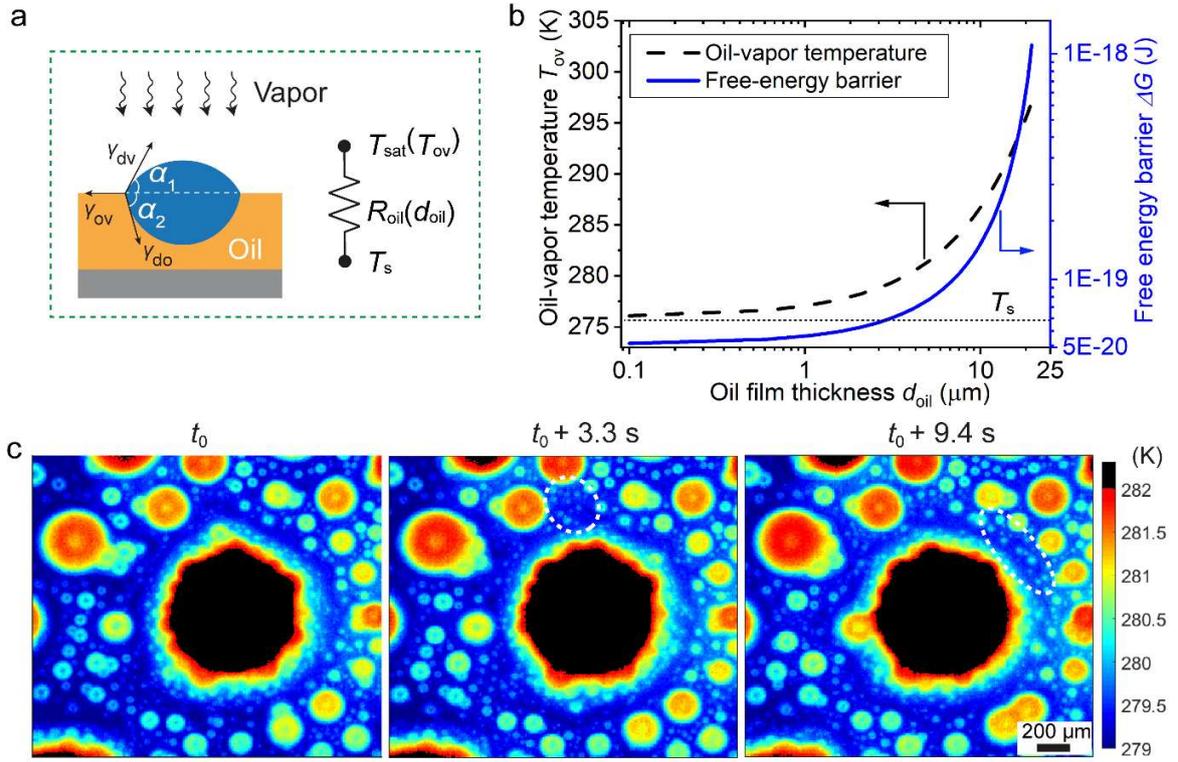

*Figure 4* Thermodynamic analysis of the nucleation energy barrier and thermal imaging of LISs during condensation. (a) The schematic shows the thermal path through the oil film with the conduction resistance $R_{oil}$. (b) Dependence of oil-vapor interfacial temperature $T_{ov}$ (black dashed line) and free energy $\Delta G$ (blue solid line) on the oil film thickness $d_{oil}$. The calculation is based on the assumption of $T_s = 276$ K, $T_v = 303$ K and a heat flux $q'' = 100$ kW/m$^2$. (c) Infrared image sequences in top-view, showing the temperature distribution at the oil-vapor interface during condensation. The post-nucleation microdroplets are highlighted by white dashed ellipses.

To experimentally validate the gradient in estimated interfacial temperatures, we measured the temperature at the oil-vapor interface from top-view during condensation using an infrared (IR) camera (Telops FAST M3k) with a 4× IR macro lens. The sample is mounted on a Linkam PE120 cold stage using double-sided thermally conductive tape. The temperature of the large droplets is shown as saturated to focus our attention on the temperature distribution of the oil. The measurement error induced by the curvature of small microdroplets and oil ridges is expected to be smaller than 1 K and neglected in this analysis.[47,48] Furthermore, the temperature readout has not been calibrated with respect to the oil's optical properties (emissivity and optical thickness) and serves as a relative comparison of interfacial temperatures only. Heat flows from the hot vapor-saturated gas to the cold substrate, leading to higher temperatures in the more elevated regions (droplets or oil-rich region), as shown in **Figure 4c**. As expected, the larger the droplet, the higher its temperature is. Interestingly, an obvious temperature gradient also establishes between oil-rich regions surrounding the droplets and oil-poor regions due to the differences in oil film thickness, which qualitatively agrees well with the theory. A large number of post-nucleation microdroplets (highlighted by white dashed ellipses) becomes visible in the freshly formed oil-poor regions at $t = t_0 + 3.3$ s and 9.4 s.

**Continuous apparent nucleation rate density on LISs**

A high droplet nucleation rate density is key to enhancing heat transfer, since droplets smaller than 10 μm account for nearly 75% of the total heat transfer.[21] The average area-availability for nucleation on LISs is expected to be smaller than on a hydrophobic surface due to the existence of oil-rich regions (oil menisci) in addition to the condensate droplets. On the other hand, the oil film continuously redistributes due to microdroplet self-propulsion, potentially leading to a higher nucleation frequency. It is interesting to examine the transient nucleation rate density within a full sweeping cycle during condensation on LISs and hydrophobic surfaces. We first examined the apparent nucleation rate density during a full natural sweeping cycle at the center of a vertically oriented LIS with Krytox GPL 104 at $T_v \approx 306 \pm 2$ K and $T_s \approx 276$ K. **Figure 5a** shows that the apparent nucleation rate density evolves with time during a sweeping cycle (also see **Supplementary video S1**). Immediately after the surface is refreshed by a large sweeping droplet ($t = 0$), the nucleation rate density increases sharply as droplets appear everywhere in the freshly cleared surface. After reaching a maximum value of $1.1 \times 10^9$ #/m$^2$s at $t \approx 12$ s, the nucleation rate density slowly decreases due to continuous accumulation of condensate droplets on the surface, which can be considered as a quasi-steady state. Finally, the entire area is swept again by a large droplet sliding down the surface due to gravity, and then the cycle begins from the top. During quasi-steady state, local coalescence events can boost the apparent nucleation rate density. At around $t = 120$ s, for example, a transient sharp increase in droplet nucleation is triggered by two large droplets coalescing (the smaller one gets "swallowed" by its bigger neighbor). Overall, the nucleation rate on the LIS is able to remain high at around $5 \times 10^8$ #/m²s at all times. If we approximate the value of nucleation rate density (NRD, #/(m²·s)) as the number of droplets with radii in the interval $r \sim [0.56, 0.93]$ μm on a per-unit condensing area with (#/m²), then the volumetric time-averaged droplet distribution density can be described as $n(r) = \text{NRD}/0.74$ (#/m³), which is in line with what others have reported for the number density of the smallest microdroplets on LISs using macroscopic measurements.[21,49] For dropwise condensation on the hydrophobic surface, however, the nucleation rate density is intermittent due to random coalescence events that suddenly created a large area of fresh nucleation sites. Then the nucleation rate locally remains at an extremely low level, until another coalescence and explosive re-nucleation occurs, as shown in the right y-axis of **Figure 5a**.

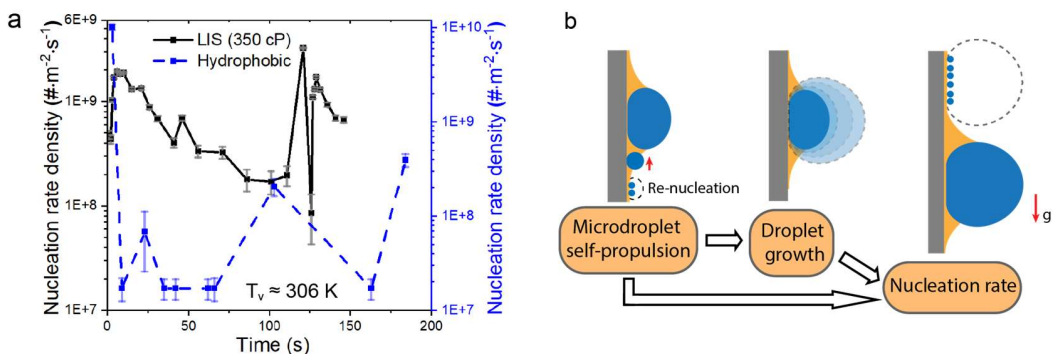

**Figure 5** *(a) Time evolution of the apparent nucleation rate density on a vertically placed LIS with Krytox GPL 104 and a hydrophobic surface at $T_v \approx 306 \pm 2$ K and $T_s \approx 276$ K (also see Supplementary video S1). (b) Schematic of the interrelationship between microdroplet self-propulsion, droplet growth, and nucleation rate during condensation on LISs.*

We propose that microdroplet self-propulsion contributes to the continuously high nucleation rate density on LISs through two mechanisms. First, small droplets (< 15 μm) continuously self-propel to oil-rich regions, providing space for the re-nucleation of condensate, as illustrated in **Figure 5b**. Second, the robust movement can potentially increase coalescence between larger droplets and thereby accelerate droplet growth rates. Driven by the release of surface energy from coalescences, droplets in the mid-size range (15 – 200 μm) are also highly mobile and efficiently refresh larger areas of the surfaces for re-nucleation, which will be discussed in detail in the following sections. As droplets grow faster, the frequency of gravity-induced sweeping will also increase, again leading to more frequent re-nucleation.

**Influence of lubricant viscosity on apparent nucleation rate density and water collection rates**

Given that microdroplets slide along a LIS with a velocity that is inversely proportional to the viscosity of the lubricant oil,[21,24] we expect the apparent NRD to also depend on lubricant viscosity. On a vertically placed LIS, natural sweeping rates and the maximum size to which droplets grow during a natural sweeping cycle depend on the vertical location on a substrate.[21] To exclude the uncertainties of the choice of the location of the observation window and to elucidate the influence of lubricant viscosity and vapor temperature on the microdroplet formation and mobility, we conducted experiments on a horizontally placed sample. We averaged the mean apparent NRD from several full artificial sweeping cycles at the respective experimental settings and plotted them in **Figure 6a**. Data from a vertically placed solid hydrophobic surface is also shown for comparison. The graph in **Figure 6a** shows that the apparent NRD increases with vapor temperature for all lubricant viscosities, as expected, since the energy barrier for heterogeneous nucleation on the oil-vapor interface decreases with the degree of subcooling according to the classical nucleation theory.[50] Importantly, we also observe that the apparent NRD on LISs increases with decreasing viscosity. We compared nucleation rate densities between horizontally (randomly selected locations) versus vertically (central location on the sample) oriented LISs and found negligible difference (see **Supplemental Information, Section S5**). As an independent approach to support our micro-scale measurements, as well as to evaluate the influence of substrate properties on the overall condensation behavior, water harvesting experiments were conducted on LISs with Krytox 102 and 106 and the hydrophobic surface at vapor temperatures of 303 K and 328 K. The results are presented in **Figure 6b**. As seen in **Figure 6b**, the LIS with Krytox 102 (lower viscosity) also has the highest water collection rate in comparison to the LIS with Krytox 106 (higher viscosity) and the hydrophobic surface. These results well support our previous claims and indicate that the NRD has a strong influence on the overall condensation performance, which should be carefully considered in modeling dropwise condensation on LISs.

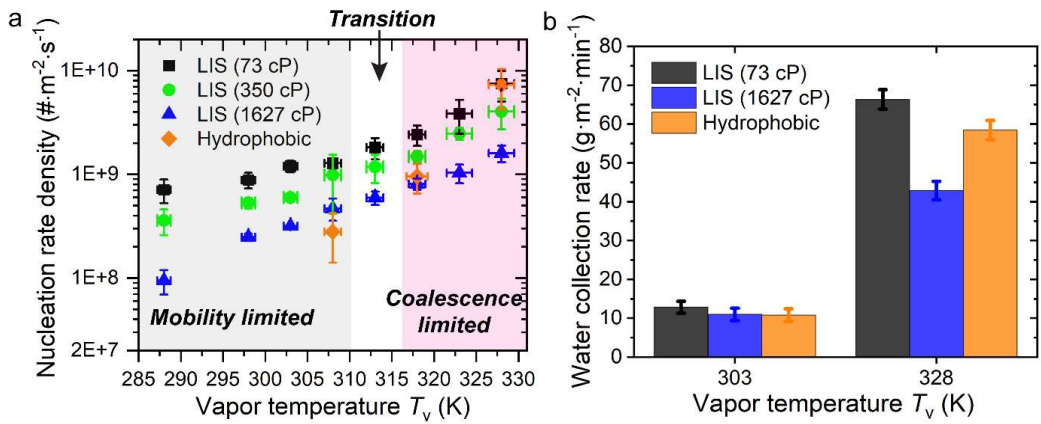

*Figure 6* (a) Dependence of apparent nucleation rate density on lubricant viscosity at different vapor temperatures $T_v$. Data was obtained from steady-state condensation experiments on horizontally placed LISs. The rate density of the newly formed 1-μm droplets on vertically placed solid hydrophobic surfaces at select $T_v$ are also presented as a benchmark. The substrate temperature was maintained at $T_s \approx 276$ K. The horizontal error bars are mainly from the fluctuation of vapor temperatures and the vertical error bars were determined from one standard error in the mean from experiments. (b) Water collection rates during condensation on hydrophobic surface and LISs with Krytox 102 and 106 at vapor temperatures of 303 K and 328 K, respectively.

We attribute the enhancement for lower viscosities to the higher mobility of condensate droplets and the resulting higher refreshing rate. As shown in **Figure 7a** for a LIS with Krytox 102, condensate droplets (outlined by dashed cyan circles) typically move long distances, as indicated by dashed lines. For a LIS with Krytox 106 (**Figure 7b**), on the contrary, the sliding velocity is lower and the travel distances of moving droplets are much shorter (also see **Supplementary video S2**). Coalescence between microdroplets of comparable size releases considerable excess surface energy, which can then contribute to additional droplet propulsion and trigger subsequent coalescence events, as observed for LISs with Krytox 102. On LISs with Krytox 106, however, merged droplets move less after coalescence due to the strong viscous resistance of the oil menisci. The coalescence between two droplets can create surplus oil when two wetting ridges are converted to one on LISs with excess oil. Since the time scale for nucleation is much faster than the time scale of oil flow (*i.e.*, of relaxation), vapor is able to nucleate in the transient oil-poor regions. Over time, the surplus oil relaxes and replenishes the oil-poor regions by forming oil ridges surrounding the newly growing droplets. More information on the nucleation and dynamics of surplus oils is available in **Section S6** of the **Supplemental Information**.

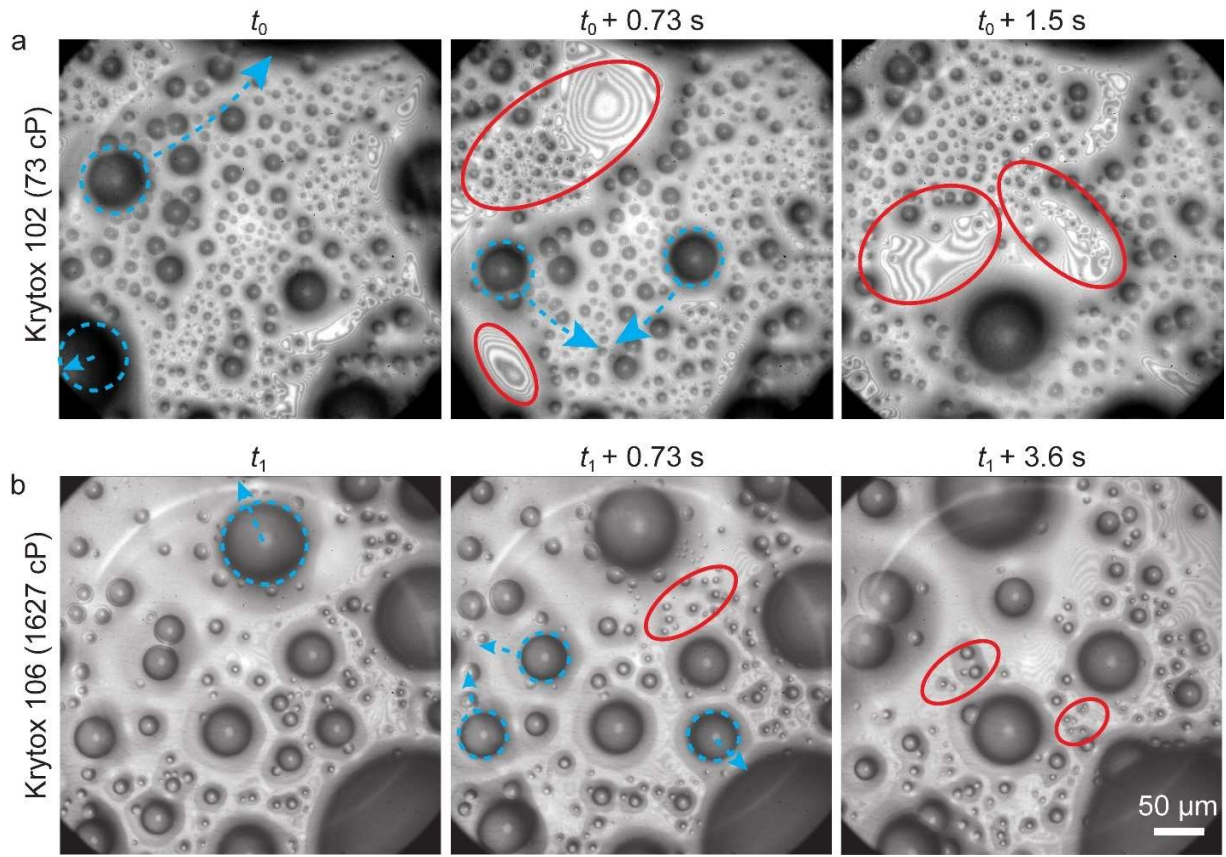

*Figure 7 Dependence of droplet mobility on lubricant viscosity. Image sequences of condensation on LISs with (a) Krytox 102 and (b) Krytox 106 at $T_v \approx 314 \pm 2$ K and $T_s \approx 276$ K. The initial times $t_0$ and $t_1$ are randomly chosen at the beginning of the $6^{th}$ sweeping cycle. Cyan colored arrows indicate the movement and direction of droplets, and the red ellipses represent the oil-poor regions with re-nucleation.*

**Characterizing microdroplet mobility**

To better compare the mobility of microdroplets on LISs with different oil viscosity, in **Figure 8** we statistically quantify the movement distances and effective frequencies of microdroplets with diameters of 15 – 200 μm on LISs with Krytox 102 and 106. Droplets commonly grow fast during movement on the LIS with the low-viscosity oil and can be characterized by long-distance movement based on the change of droplet sizes. **Figure 8a** clearly shows that the microdroplets on the LIS with the lower viscosity cover longer distances than on a high-viscosity LIS for the entire range of microdroplet sizes. Within the same condensation time and area, the number of droplet movements (for $D > 40$ μm) is about twice on the LIS with Krytox 102 than with Krytox 106, as shown in **Figure 8b**. The frequent long-distance movement of mid-sized microdroplets can effectively and continuously sweep other microdroplets occupying the surface and leave fresh oil-poor regions behind their trajectories (highlighted by red ellipses in **Figure 7a**). Subsequently, a high density of emerging droplets is observed. This continuous and gravity-independent microdroplet sliding not only contributes to the fast growth of the moving droplets and their coalescence partners, but also frequently exposes colder regions previously occupied by the big droplets to hot vapor, enabling a higher re-nucleation frequency. The movement of droplets with a diameter over 200 μm is also observed on LISs, which refreshes large areas for nucleation (similar to hydrophobic surfaces). However, the ratio of movement length to the droplet diameter and movement frequency are much lower than for smaller droplets, and is thus not included in **Figure 8**.

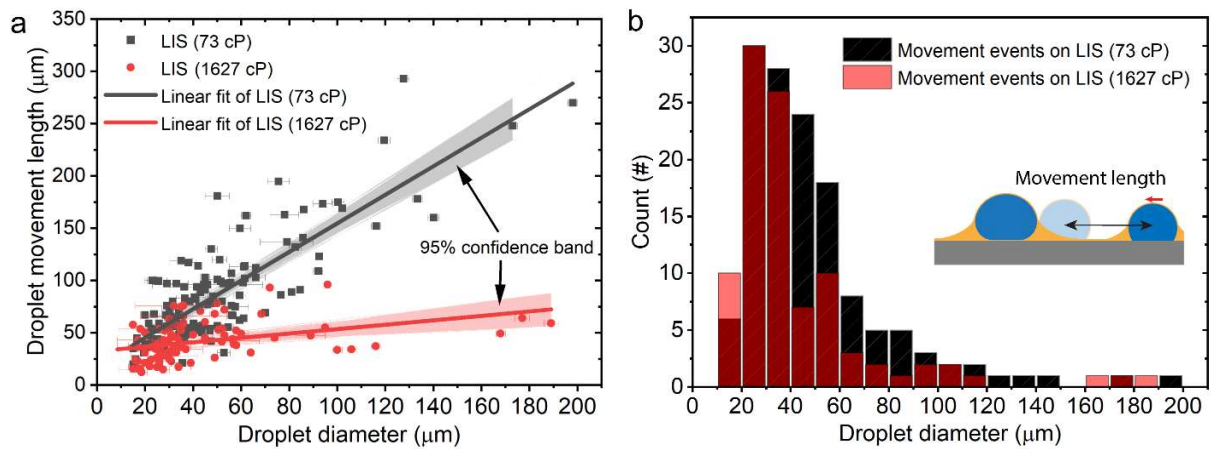

***Figure 8*** *Statistical analysis of microdroplet (15 – 200 μm) movement on LISs with Krytox 102 (73 Cp) and 106 (1627 cP) at a vapor temperature of 312 K. (a) Dependence of microdroplet movement on droplet size. The horizontal error bars derive from droplet growth during the movement. (b) Frequency distribution of droplet movement for different droplet diameters during a 1-minute condensation interval within an area of 0.117 mm².*

**Outperformance of the hydrophobic surface over LISs with high viscosity oil at high vapor temperatures**

Compared to the very transient nucleation behavior with sporadic nucleation bursts on hydrophobic surfaces, LISs achieve a more uniform nucleation rate density due to high microdroplet mobility. Nevertheless, if the oil viscosity of the LISs is considerably high, one can expect the microdroplet mobility to severely decrease and the characteristic coalescence time to significantly increase due to delayed oil film drainage between the droplets. Here, we compare the apparent NRD and overall water collection rate between LISs and hydrophobic surfaces over a wide range of vapor-substrate temperature differences $\Delta T = T_v - T_s$. Generally, as shown in **Figure 6**, LISs can achieve higher nucleation and water collection rates than a solid hydrophobic surface at low vapor temperatures ($T_v < 312$ K). However, hydrophobic surfaces start to outperform LISs with high viscosity oils (1627 cP, and to a certain extent even 350 cP) when $T_v > 317$ K. For example, at $T_v = 328$ K, the water collection rate reaches 58 g·mm$^{-2}$·min$^{-1}$ on the hydrophobic surface; nearly 30% higher than that of LISs with Krytox 106. This indicates that condensation on LISs might not be as superior as commonly assumed for high degrees of subcooling $\Delta T$. In the following, we investigate the mechanisms underlying different condensation performances of LISs compared to the solid hydrophobic surface at low and high vapor temperatures separately.

At lower temperatures, the kinetics of droplet nucleation and growth and microdroplet mobility are rate-limiting. As discussed above, droplets intrinsically nucleate more easily on LISs (independent of lubricant viscosity) than on solid hydrophobic surfaces ascribed to a lower Gibb's energy barrier,[16] which can contribute to a higher NRD on LISs than on the hydrophobic surface. However, more importantly, small emerging microdroplets vigorously self-propel to the oil-rich regions, allowing for continuous re-nucleation. This movement is absent on hydrophobic surfaces.

As shown in **Figure 6a,** a 3× increase in the apparent NRD can be achieved by lowering the lubricant viscosity from 1627 cP (Krytox GPL 106) to 73 cP (Krytox GPL 102), whereas the enhancement (presumably due to the lower nucleation energy barrier) of the LIS with Krytox 106 compared to the hydrophobic surface is only 1.5×. This indicates that droplet mobility plays a dominant role for enhancing the NRD at low vapor temperatures.

For condensation at higher $\Delta T$, the nucleation density is high and the time scale for individual droplet growth through direct condensation is small, meaning that high droplet mobility is becoming increasingly unimportant as compared to droplet coalescence. An oil with higher viscosity has a longer characteristic coalescence time scale due to slow oil drainage in-between approaching water droplets, which becomes the rate-limiting factor at high vapor temperatures.[27] Thus, we observe a transition from mobility-limited nucleation rates, where LISs of all viscosities outperform a solid hydrophobic surface, to coalescence-limited nucleation at higher degrees of subcooling, where the solid hydrophobic surface becomes increasingly more efficient (gas has a negligible drainage resistance). To substantiate this hypothesis, we characterize the growth rate of microdroplets on LISs with Krytox 102 and 106 oils and hydrophobic surfaces at two typical vapor temperatures ($T_v \approx 302$ K and 327 K) in the following section.

**Enhanced droplet growth rate based on higher microdroplet mobility**

Five droplets are randomly selected on fresh LISs and hydrophobic samples at each vapor temperature and viscosity. The time evolution of the droplet diameters is tracked and plotted in **Figure 9**, where $t = 0$ denotes the moment when droplets first become visible. For dropwise condensation on hydrophobic surfaces, the droplet growth rate has traditionally been described by a power law,[19,41]

$$r(t) = at^b, \qquad (2)$$

where $a$ is fitting number, and $b$ is the growth exponent constant. In eq. (2), two growth mechanisms – diffusion-driven (*i.e.*, direct condensation) and coalescence-driven growth – are distinguished by using different values of $b_1$ and $b_2$, with an effective transition radius $r_e$. In the diffusion-driven regime ($r < r_e$), droplet growth is nominally limited by diffusion of water vapor from the surroundings to the droplet, and the value $b$ should be comparable for both LISs and the hydrophobic surface. Once the droplet diameter reaches $r_e$, droplet growth is enhanced by coalescence between droplets, and the growth rate of microdroplets on LISs and the hydrophobic surface starts to diverge. The values listed in the chart of **Figure 9**, based on experimental data fitting, show that $b$ is comparable on all surfaces at the low vapor temperature of 302 K. However, we observe that the value of $r_e$ is smaller on LISs than on the hydrophobic surface. Especially when condensate droplets are sparse at the low vapor temperature, oil menisci surrounding droplets can act as bridges to facilitate droplet coalescences and lower $r_e$. Droplets enter the coalescence-dominated growth regime at $r_e \approx 2.5$ μm on LISs, whereas it is at $r_e \approx 9$ μm for droplets on the hydrophobic surface. This difference exerts significant effects on the overall droplet growth. For example, before reaching a diameter of 75 μm, a typical droplet coalesces 40 times on a LIS with Krytox GPL 102, whereas only 15 times on the hydrophobic surface. Interestingly, when the vapor temperature increases to 327 K, $b_{2,\text{LIS}}$ increases slightly, but $b_{2,\text{hydro}}$ increases significantly and even surpasses $b_{2,\text{LIS}}$[106]. We can foresee that droplets will reach a diameter of 150 μm faster on hydrophobic surfaces than on the high-viscosity LIS at $T_v = 327$ K. At the large vapor-substrate temperature difference, the diffusion rate and droplet density on the surface are large (*i.e.*, droplet-droplet distance is small). We hypothesize that self-propulsion becomes less important for the microdroplet growth since the droplets do not need to travel a long distance to coalesce with others. At the same time, as discussed above, on LISs, droplet coalescence can be delayed due to the necessity of oil film drainage between the droplets. Especially higher-viscosity lubricants will be coalescence-limited for these conditions.

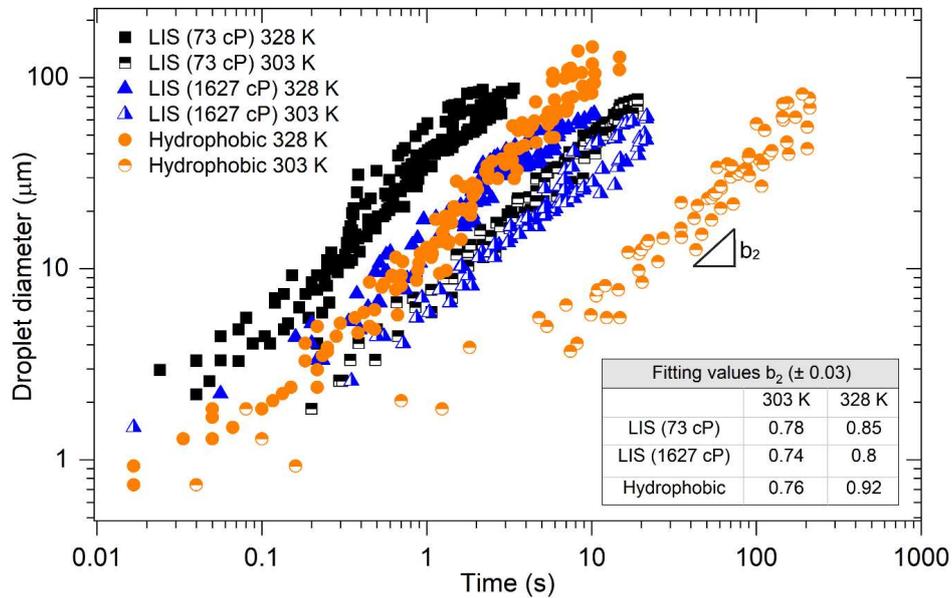

*Figure 9* Droplet growth on lubricant-infused surfaces and hydrophobic surfaces during condensation on hydrophobic surfaces and LISs with Krytox 102 and 106 at vapor temperatures of $T_v = 303$ K and 328 K. The inserted table lists fitting values $b_2$ for each case.

**Influence of lubricant loss on NRD and water collection rate**

Continuous lubricant drainage and loss during condensation has been one of the main concerns for the practical implementation of LISs in commercial thermal management applications, and the oil depletion mechanisms have been extensively studied.[35,51–54] We assume that the loss of lubricant on LISs has complex and important effects on the NRD. Intuitively, an overall thinner oil film due to oil depletion will result in a lower thermal resistance and lower oil-vapor interfacial temperature, decreasing the nucleation energy barrier and increasing the nucleation rate. However, we cannot ignore the crucial negative aspect of smaller oil ridges that come along with a thinner oil film: lower mobility of droplets. Microdroplets robustly self-propel long distances due to overlapping menisci between droplets.[28] The size of oil menisci (or wetting ridges) is significantly smaller for a starved regime than that on LISs with excess oil.[55] Smaller-sized oil menisci will significantly lower the probability of building wetting 'bridges', *i.e.*, creating overlapping menisci, between droplets. Droplets would behave as isolated islands and would not coalesce with neighboring droplets until they were much closer (*i.e.*, physically touching), similar to the behavior on a solid hydrophobic surface. In this work, we attribute the higher NRD on LISs with low-viscosity oil mainly to its higher droplet mobility and coalescence rates. Due to the competition of droplet mobility and thermal resistance, we expect that there should be an optimal oil film thickness to achieve the best performance. Although there is no full depletion (some oil is always left due to strong capillary forces in the nanoporous substrate),[53] we expect that the NRD will decrease and approach equilibrium with the loss of lubricant. Although LISs with low-viscosity oil can achieve high NRD, the lubricant depletion rate will be faster on these surfaces as compared to the higher viscosity oil infused surfaces. Hence, lubricant loss must be carefully considered for achieving an average high NRD in longtime condensation applications.

## Conclusions

To summarize, we qualitatively showed that nucleation has a spatial preference on the microscopically uneven oil film during condensation on LISs. Using high-speed high-resolution IR imaging, we measured an up to 3 K lower oil-vapor interface temperature of the oil-poor compared to the oil-rich regions, which decreases the nucleation barrier energy by up to an order of magnitude. Since the oil film continuously redistributes due to robust microdroplet self-propulsion, an uninterrupted high nucleation rate was found on LISs (in contrast to solid hydrophobic surfaces, where nucleation occurs quite erratically). We examined the influence of oil viscosity on nucleation, growth, and heat transfer (represented by average water collection rates) for a wide range of vapor-substrate temperature differences. The results showed the nucleation rate density is proportional to the temperature difference, in agreement with the classical nucleation theory. Remarkably, at a given temperature difference, a higher apparent nucleation rate density can be achieved by lowering the lubricant viscosity. This increase can be attributed to higher mobility and growth rate of microdroplets on LISs with the low-viscosity Krytox oil. We quantitatively demonstrated that a higher frequency and longer length of microdroplet movement on low-viscosity oil (Krytox 102) enhances nucleation. Small microdroplets spontaneously move towards larger microdroplets in oil-rich regions, leaving behind a fresh, cold surface for re-nucleation. The robust movement of relatively large microdroplets can also efficiently refresh the surface by sweeping all droplets in their trajectories, again leaving an empty oil-poor region behind for re-nucleation. At low vapor-substrate temperatures, LISs of all lubricant viscosities (73 cP – 1627 cP) showed higher apparent nucleation rate densities and water collection rates than the hydrophobic benchmark surface. At these low degrees of subcooling, the nucleation density is small (*i.e.*, droplet-droplet distance is large) and the high droplet mobility and self-propulsion on LISs greatly enhance droplet growth ("mobility limited"). However, LISs with high viscosity oils (350 cP and 1627 cP) unexpectedly underperformed at larger temperature differences compared to the solid hydrophobic surface because of longer characteristic coalescence time scales associated with oil drainage. At these high degrees of subcooling, the inter-droplet spacing is minimal and direct condensation growth rates are naturally large, shifting the efficiency of droplet growth to fast coalescence ("coalescence limited"). Thus, interestingly and initially counter-intuitively, solid hydrophobic surfaces are able to achieve up to 30% higher water collection rates than LISs with a high lubricant viscosity (*e.g.*, Krytox 106) at these higher vapor temperatures. These findings will be helpful to comprehensively understand the interplay of the lubricant film and nucleation and to enhance apparent nucleation and water collection rates by tailoring the lubricant viscosity.

## Associated Content

**Supporting Information**

The Supporting Information is available free of charge.

Initial oil film thickness; Image of experimental setup; Conduction resistance determines nucleation preference; Dependence of free-energy barrier $\Delta G$ on oil film thickness; Comparison of apparent NRD between horizontally versus vertically oriented samples; Nucleation and dynamics of surplus oil replenishment in oil-poor regions after droplet coalescence (PDF)

Video S1 Time evolution of nucleation rate density on a vertically place lubricant-infused surface (Fig. 6a from t=0 sec to t=80 sec) (MP4)

Video S2 Droplet mobility on LISs with different viscosity oils (MP4)

## Author Information

**Corresponding Author**

**Patricia Weisensee** – Department of Mechanical Engineering & Materials Science, Washington University in St. Louis, Saint louis, Missouri 63130, United States; Institute of Materials Science and Engineering, Saint louis, Missouri 63130, United States; orcid.org/0000-0003-0283-9347; Email: p.weisensee@wustl.edu

Authors

**Jianxing Sun** – Department of Mechanical Engineering & Materials Science, Washington University in St. Louis, Saint louis, Missouri 63130, United States; orcid.org/ 0000-0003-3784-9908

**Xinyu Jiang** – Department of Mechanical Engineering & Materials Science, Washington University in St. Louis, Saint louis, Missouri 63130, United States

## Conflicts of Interest

We report no conflicts of interest.

## Acknowledgement

The authors would like to thank Michael Lorberg for assisting with droplet counting, and Jonathan Boreyko for elucidating discussions. This material is based upon work supported by the National Science Foundation under Grant No. 1856722. The authors acknowledge the use of instruments and staff assistance from the Institute of Materials Science and Engineering (IMSE), and the Jens Environmental Analysis Facility at Washington University in St. Louis.

# Supplemental Information

# Enhanced Water Nucleation and Growth Rate Based on Microdroplet Mobility on Lubricant-Infused Surfaces

Jianxing Sun[a], Xinyu Jiang[a], and Patricia B. Weisensee[a,b,*]

a. Department of Mechanical Engineering & Materials Science, Washington University in St. Louis, USA

b. Institute of Materials Science and Engineering, Washington University in St. Louis, USA

* corresponding author: p.weisensee@wustl.edu

**Section S1 Initial oil film thickness**

**Section S2 Image of the experimental setup**

**Section S3 Conduction resistance determines nucleation preference**

**Section S4 Dependence of free-energy barrier *ΔG* on oil film thickness**

**Section S5 Comparison of apparent NRDs between horizontally versus vertically oriented samples**

**Section S6 Nucleation and dynamics of surplus oil replenishment in oil-poor regions after droplet coalescence**

**Section S1 Initial oil film thickness**

The initial oil film thickness was measured using high-speed interferometry. **Figure S1a** shows a fresh LIS sample where a submerged condensate droplet is present on the sample just prior to protruding the oil-vapor interface, highlighting the uniform oil film thickness of a fresh sample. The wide-spaced interference fringes are caused by the slight tilt of the sample. Any defects or condensate droplets (as, for example, outlined with the dashed rectangle in the center of the image) protruding out of the oil film disturb the uniform interference patterns. The detailed experimental information on this method can be found in our previous work.[1] Using this analysis, we compared the oil film thickness for different spin coating speeds and lubricant viscosities, as shown in **Figure S1b**, and observe good agreement between experimental measurement results and the spin coating model ($d_{oil} \sim (\mu/t\omega^2)^{1/2}$).

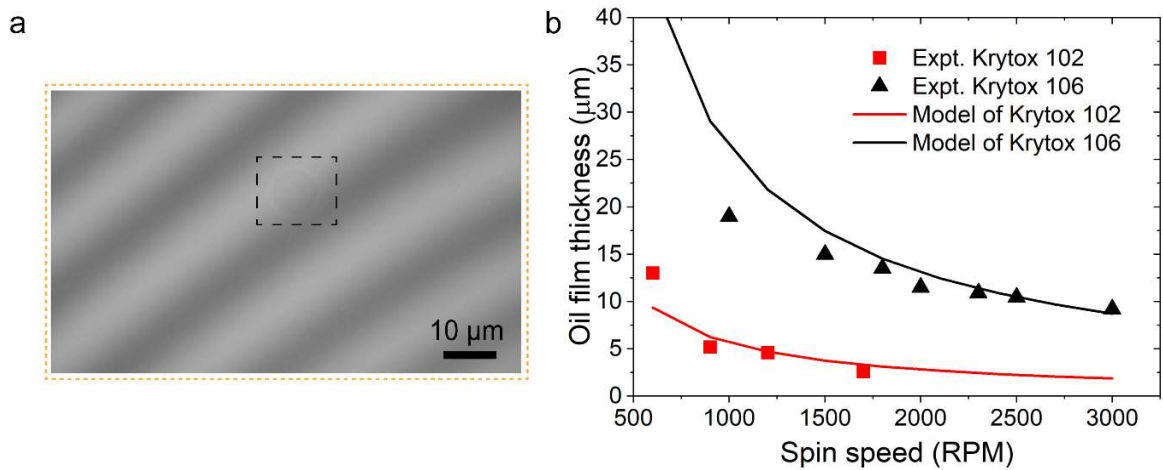

*Figure S1 (a) Interferometry image of a fresh LIS sample at the start of gentle condensation. The widely spaced interference patterns are caused by a slight tilt of the sample. A condensed microdroplet protruding the oil-vapor interface appears in the center of the dashed rectangle. (b) Relationship between oil film thickness and RPM, comparing experimental measurements and the spin coating model, $d_{oil} \sim (\mu/t\omega^2)^{1/2}$*

**Section S2 Image of the experimental setup**

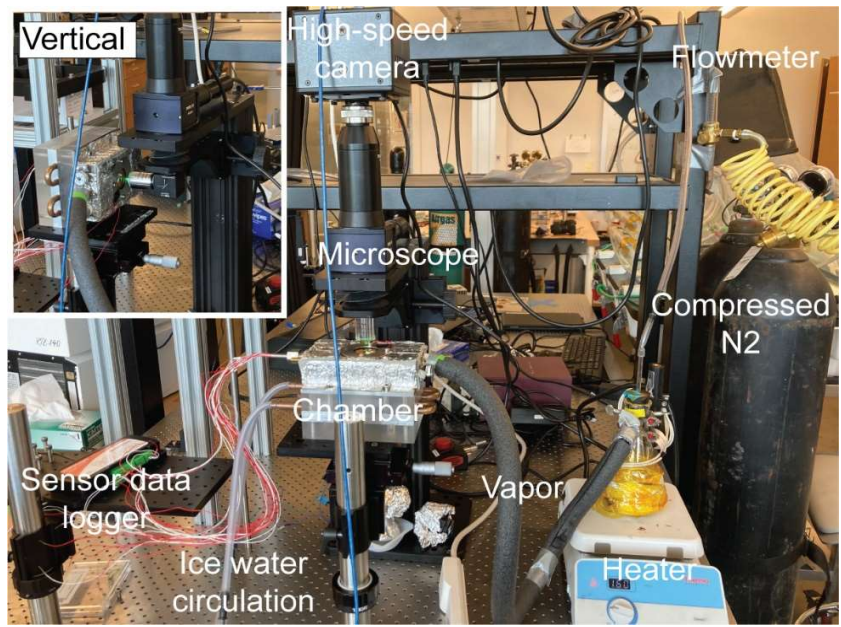

*Figure S2 Image of the experimental setup with a horizontally placed chamber. Insert: Vertically placed chamber, using a 90° mirror cage between the microscope wheel and the objective lens.*

**Section S3 Conduction resistance determines nucleation preference**

The dry-zones during condensation on LISs in the presence of NCGs spatially overlap with oil menisci, *i.e.*, oil-rich regions. Both a low vapor concentration (due to said dry-zones) close to larger droplets and a high conduction resistance of the thick oil film make oil-rich regions unfavorable for nucleation. To elucidate which of these factors dominates the spatial preference for nucleation, we investigate nucleation dynamics nearby similar-sized droplets on a LIS with Krytox 106 at two vapor temperatures (~ dry-zone dimension $\delta_{dz}$) for two very different meniscus sizes ($l_m$). Then, we analyze the distance of the nearest location of nucleation away from the edge of the large droplet. Panels A1 and A2 ($T_v \approx 297$ K) of **Figure S3** show a locally starved oil film, leading to the formation a small oil-rich region with a width of $l_m \approx 10$ μm (marked with the yellow dashed line) surrounding a large droplet ($D \approx 182$ μm; marked with the white dotted line). The dry-zone width $\delta_{dz}$ has previously been reported to be 15 – 20 μm at a subcooling of 35 K and is hence larger than the meniscus in this example.[2] We observe that droplets successfully nucleate at the boundary of oil-rich and oil-poor regions; within the dry-zone. In **Figure S3**, panels B1 and

B2 ($T_v \approx 317$ K), we examine the nucleation locations for larger oil-rich regions ($l_m \approx 30$ μm, $D \approx$ 167 μm). At this higher temperature, we expect $\delta_{dz} < 10$ μm, thus the oil-rich/poor transition is well outside the dry-zone. We observe that the nearest point of nucleation is approximately 33 μm away from the central droplet edge, just outside the oil-rich region and far away from the dry-zone. Based on these observations, we can conclude that the oil film thickness, not the existence of dry-zones, determines the spatial preference for nucleation.

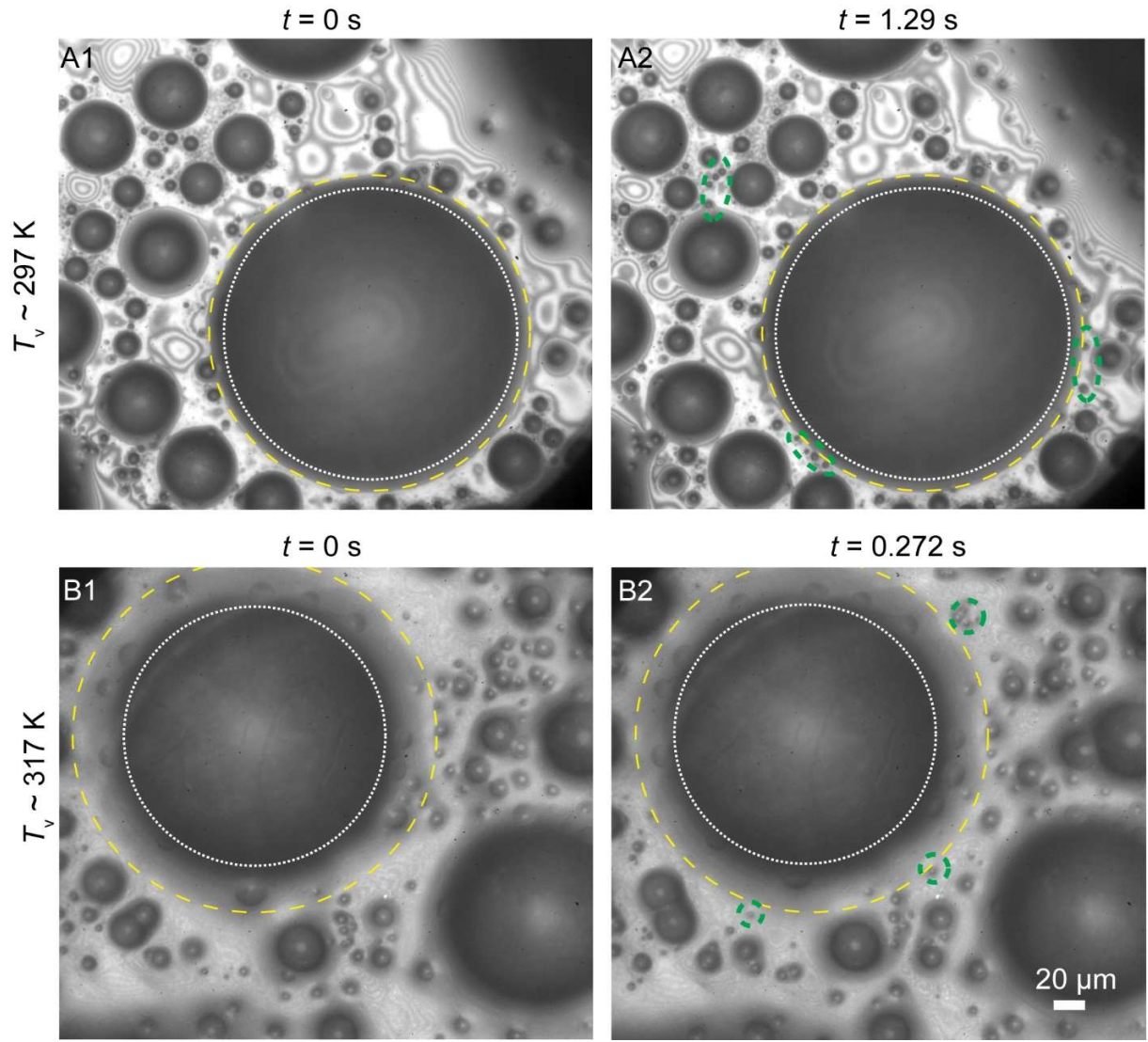

*Figure S3 Nucleation dynamics nearby similar-sized droplets on a LIS with Krytox 106 at two vapor temperatures with $T_s \approx 276$ K and for two very different meniscus sizes. The boundary between oil-rich and oil-poor regions is marked with the yellow dashed line and the droplet profile is marked with the white dotted line. New droplets are highlighted with green dashed lines.*

**Section S4 Dependence of free-energy barrier $\Delta G$ on oil film thickness**

According to the classical nucleation theory (CNT), water vapor molecules diffuse through the vapor boundary layer to the oil-vapor interface and then form clusters which can grow only if their sizes exceed a critical size $r_c$.[3] The free-energy barrier $\Delta G$ corresponding to the critical nucleation radius $r_c$ is:[4]

$$\Delta G = \gamma_{dv} A_{dv} + \gamma_{do} A_{do} - \gamma_{ov} \pi r_c^2 - (P_d - P_v) V_d. \tag{S1}$$

In eq. (S1), $V_d$ is the volume of the nucleus, and $A_{i,j}$ are the interfacial areas between phases i and j, with the subscripts o, d, v, ov, do, and dv representing the oil, droplet, vapor, oil-vapor, droplet-oil, and droplet-vapor phases, respectively. $(P_d - P_v)$ is the Laplace pressure difference between vapor and droplet. As illustrated in the schematic of **Figure 4a** in the manuscript, water vapor nucleates at the oil-vapor interface and the droplet can be approximated as two hemispherical caps with a deformation angle $\alpha_1$ of the upper hemispherical cap and angle $\alpha_2$ of the lower hemispherical cap. Hence, the interfacial areas $A_{dv}$, $A_{do}$, and the volume $V_d$ can be described by:[3,5]

$$A_{dv} = 2\pi R_1^2 (1 - \cos \alpha_1), \quad A_{do} = 2\pi R_2^2 (1 - \cos \alpha_2), \tag{S2}$$

$$V_d = \frac{\pi R_1^3}{3}[2 - \cos\alpha_1(2 + \sin^2\alpha_1)] + \frac{\pi R_2^3}{3}[2 - \cos\alpha_2(2 + \sin^2\alpha_2)], \tag{S3}$$

where $R_1$ and $R_2$ are the equilibrium radii of upper and lower curvatures, respectively, and can be obtained through $R_1 \sin\alpha_1 = R_2 \sin\alpha_2 = r_c$. The critical nucleation radius and the Laplace pressure difference $(P_d - P_v)$ can be estimated, respectively, using [6]:

$$r_c = \frac{2\gamma_{ov}}{(P_v - P_d)} \frac{\sin\alpha_1 \sin\alpha_2}{\sin(\alpha_1 + \alpha_2)}, \tag{S4}$$

$$P_d - P_v = \frac{kT_{ov}}{V_{m,d}} \ln\left(\frac{P_v}{P_{sat}}\right), \tag{S5}$$

where $P_{sat}$ and $V_{m,d}$ represent the saturation pressure and liquid molar volume at the saturation pressure, respectively.

Since Krytox oils cloak water in our three-phase (oil-water-vapor) system, the Neumann triangle law collapses. To calculate the angles $\alpha_1$ and $\alpha_2$, we modify the Neumann triangle equations by replacing the interfacial tension $\gamma_{dv}$ with a combined effective surface tension $\gamma_{eff}$ [7,8]:

$$\cos\alpha_1 = \cos(\pi - \theta_v) = \cos\theta_v = \frac{\gamma_{do}^2 - \gamma_{eff}^2 - \gamma_{ov}^2}{2\gamma_{ov}\gamma_{eff}}, \tag{S6-1}$$

$$\cos\alpha_2 = \cos\theta_o = \frac{\gamma_{eff}^2 - \gamma_{ov}^2 - \gamma_{do}^2}{2\gamma_{do}\gamma_{ov}}. \tag{S6-2}$$

With a measured value of $\theta_v \approx 150°$ for the same material system,[1] the value of $\gamma_{eff}$ can be determined by eq. (S6-1) to $\gamma_{eff} \approx 66$ mN/m. By combining eqs. (S1) – (S6) and eq. (1) of the manuscript, we can obtain the change of the free-energy barrier $\Delta G$ versus the oil film thickness $d_{oil}$ (from oil-poor regions to oil-rich regions on the LIS), corresponding to different temperatures $T_{ov}$, as seen in **Figure 4b** of the manuscript.

It is worthy to note that this thermal conduction and nucleation energy barrier analysis might not be applicable to the part of oil-rich regions very close to big droplets where oil-immersed microdroplets and strong latent heat release from condensation near the apparent droplet-oil-vapor triple line might exist. To include all of these factors is nontrivial and would require a full 3D, transient analysis (likely numerical). Along with the high thermal conductive resistance (thicker oil film), the latent heat released from condensing vapor on droplets will increase the temperature of the droplet-vapor or oil-vapor interface in oil rich regions to a certain extent, making these locations less favorable for nucleation. Due to the strong temperature gradients and a thin film approximation between vapor and substrate, we neglected possible mobility-induced convection heat transfer in the oil film. Our model thus represents a qualitative estimate rather than a quantitative prediction of interfacial temperatures.

**Section S5 Comparison of apparent NRDs between horizontally and vertically oriented samples**

By averaging the values within a natural sweeping cycle of Figure 5a (vertically placed sample), we obtain an average nucleation rate density of about $6.07 \times 10^8$ #/m²s. The data point from Figure 6a (horizontally placed samples) with the same lubricant viscosity and temperature difference, but in a horizontal orientation, shows a nucleation rate density within a sweeping cycle of approximately $5.98 \times 10^8$ #/m²s – basically the same. Although the droplets within an artificial sweeping cycle were able to grow beyond the maximum size that droplets can grow to within a natural sweeping cycle, we found the two NRD values to be comparable. To elucidate the influence of lubricant viscosity and the vapor temperature, we averaged the mean nucleation rate density from several full artificial sweeping cycles at the respective experimental settings.

**Section S6 Nucleation and dynamics of surplus oil replenishment in oil-poor regions after droplet coalescence**

The coalescence between two droplets can create excess oil regions. However, droplets, especially larger droplets, typically already have a meniscus, *i.e.*, an oil-rich region surrounding them. Upon coalescence, the two menisci converge, and the net oil-rich area decreases. Over time, the surplus oil relaxes and forms a thin uniform film. As shown in **Figure S4**, at $t$ = 4 ms, two neighboring droplets coalesce and leave behind oil footprints (menisci). Immediately thereafter, a number of small nucleated water droplets (partially marked with green circles) emerge in the regions previously occupied by larger droplets. By $t$ = 148 ms, the surplus oil fully relaxes and forms new oil wetting ridges surrounding newly growing droplets.

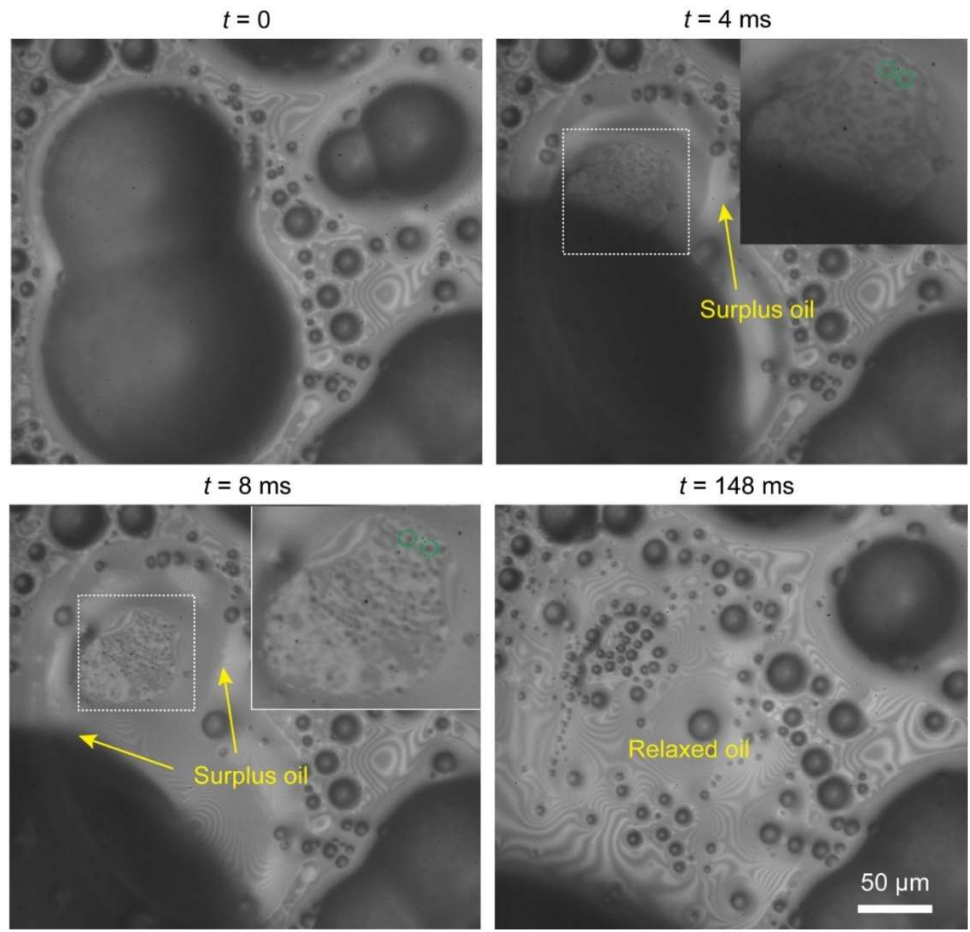

*Figure S4 Image sequences of nucleation and oil dynamics after droplet coalescence during condensation on a LIS with Krytox 106 at $T_v \approx 327 \pm 2$ K and $T_s \approx 276$ K.*